\documentclass[pdflatex,sn-mathphys-num]{sn-jnl}

\usepackage{graphicx}%
\usepackage{multirow}%
\usepackage{amsmath,amssymb,amsfonts}%
\usepackage{amsthm}%
\usepackage{mathrsfs}%
\usepackage[title]{appendix}%
\usepackage{xcolor}%
\usepackage{textcomp}%
\usepackage{manyfoot}%
\usepackage{booktabs}%
\usepackage{listings}%
\usepackage{graphicx}
\usepackage{subcaption}

\theoremstyle{thmstyleone}%

\theoremstyle{thmstyletwo}%

\theoremstyle{thmstylethree}%

\begin{document}

\title[Black Holes as Catalysts for Cosmic String Detection and Axion Dark Matter Genesis]{Black Holes as Catalysts for Cosmic String Detection and Axion Dark Matter Genesis}

\author[1]{\fnm{Ishan} \sur{Swamy}}

\author*[1]{\fnm{Deobrat} \sur{Singh}}\email{deobrat.singh@mitwpu.edu.in}

\affil[1]{\orgdiv{Department of Physics}, \orgname{Dr. Vishwanath Karad MIT-World Peace University}, \orgaddress{\street{Kothrud}, \city{Pune}, \postcode{411038}, \state{Maharashtra}, \country{India}}}

\abstract{The global $U_{PQ}(1)$ (Peccei-Quinn) symmetry, introduced to resolve the strong CP problem, predicts the existence of the axion—a pseudo-Nambu-Goldstone boson that is also a leading candidate for dark matter. The spontaneous breaking of this symmetry produces global strings, which decay by radiating massive axions and gravitational waves.In this study, we investigate the decay of cosmic axion strings around a Schwarzschild black hole, estimating the energy radiated due to string contraction as well as the decay timescales of the string loops. For primordial black holes (PBHs) with masses as small as $10^{-16} M_\odot$, the total radiated energy by the string is found to be on the order of $10^{27}$ GeV, encompassing both axion emission and gravitational waves. A key finding is that the presence of a central black hole significantly accelerates the decay of cosmic string loops, substantially reducing their lifetimes. We present these results as an initial estimate of axion radiation from PBH–cosmic string systems along with the decay time as an important observational signature for axions strings.}


\maketitle

\section{Introduction}\label{intro}
Peccei-Quinn symmetry has been proposed in the Standard Model as a potential solution to the strong CP problem \cite {peccei77}. The pseudo-goldstone boson of this model, called the axion, \cite{Weinberg78} also serves as a dark matter candidate \cite{Preskill83, Abbott83, Dine83, Berezhiani91, Khlopov99}. This $U_{PQ}(1)$ symmetry when spontaneously broken, leads to the formation of global strings called cosmic axion strings \cite{Davis86}, \cite{Vilenkin00}. Multiple studies have shown that global cosmic strings decay into radiation which can be massless goldstone bosons, and massive particles like axions \cite{Vilenkin86}, \cite{Saurabh20} and even gravitational waves (GW) \cite{Turok84, Vachaspati85, Matsunami19}. 

Research on cosmic strings also involves their interaction with black holes, which lead to interesting effects. Strings when attached to a rotating black hole have been proposed to extract the black hole's rotational energy \cite{Kinoshita16}, \cite{Xing21}. Further, string interactions with primordial black holes (PBHs) suggest the formation of a large black hole-cosmic string network \cite{Vilenkin18}. Depending on their trajectories, strings near black holes may be chaotically captured or scattered \cite{Larsen94}, \cite{Frolov99}, while circular strings can exhibit vibrational and oscillatory behavior around a Schwarzschild black hole \cite{Larsen98}, \cite{Churilova24}. Further, recent studies have derived the modified relativistic orbits for cosmic string - black hole systems \cite{parth24}. Previous studies by the authors have also explored cosmic string–black hole systems in low-mass X-ray binaries, suggesting that the influence of strings on black hole spin and the system's orbital period and accretion structure could produce observable signatures, providing potential evidence for the existence of cosmic strings \cite{Swamy25, Singh25, Swamyprd}. Recent work has also examined the role of primordial black holes (PBHs) as a potential source of axionlike particles (ALPs) through Hawking radiation \cite{Agashe23}.\\

In this manuscript, we study the dynamical evolution of a circular cosmic axion string around a black hole and estimate the energy radiated in the form of axions during the contraction of the string. We establish the equation governing the radiation and the resultant increase in energy density, and then analyse the influence of black hole on the string's lifecycle. The energy radiated by this mechanism is found to be massive, ranging from $\sim 10^{25}$ GeV up to $\sim 10^{53}$ GeV for increasing black hole mass and axion decay constant. A primary finding of this work is that the central black hole acts as a catalyst for string decay; the intense gravitational field drastically accelerates the loop's contraction, thereby shortening its lifetime compared to vacuum solutions and fundamentally altering the axion emission profile.  These preliminary findings motivate further investigation, particularly into the co-emission of gravitational waves as a promising observational channel. The detection of such axion radiation, coupled with the characteristic GW signatures of cosmic strings, offers a multi-messenger pathway to validating this mechanism and probing the existence of cosmic strings. All calculations are performed assuming natural units ($\hbar = c = 1$) throughout the manuscript.

\section{Circular string around a black hole}\label{maths}
In the axion string model, the energy density is primarily governed by the axion decay constant $f_a$ and the logarithmic divergence associated with the string's thickness $\delta$ and its characteristic length scale (cutoff radius) $\Lambda$ \cite{Davis86},
\begin{equation}
\mu = 2\pi f_a^2 \ln(\Lambda/\delta)
\label{mu}
\end{equation}

We modify the tension of general string in (\ref{mu}) to fit a circular string loop by identifying the cutoff radius $\Lambda$ with the size of the closed loop \cite{Vilenkin87},  taken as
\begin{equation} 
\Lambda = 2\pi r
\label{cutoff}
\end{equation}
with $r$ being the radius of the circular string. \\

To study the evolution of this string loop, we embed it into a Schwarzschild background such that the center of the black hole coincides with the center of the loop. Thus, the schwarzschild metric in presence of the string is given by \cite{Larsen98},
\begin{equation}
ds^2 = -(1-2GM/r)dt^2 + (1-2GM/r)^{-1}dr^2 + r^2(d\theta^2 + sin^2\theta d\phi^2)
\label{metric}
\end{equation}
where the authors have taken (by ansatz)
\begin{equation}
x^0 = t = t(\tau),\quad  x^1 = r = r(\tau),\quad x^2 = \theta = \pi/2,\quad x^3 = \phi = \sigma.
\label{ansatz}
\end{equation}
for the worldsheet coordinates $(\tau,\sigma)$ of the string. From (\ref{metric}), we get the equations of motion as 
\begin{equation} 
\dot{t} = \frac{E_m}{1-2GM/r}
\label{tdot}
\end{equation}
\begin{equation}
\dot{r}^2 = E_m^2 - r^2(1-2GM/r)
\label{rdot}
\end{equation}
where $E_m$ is the energy of the string at maximal radius. These expressions allow us to track the trajectory of the string, as they set the physical boundaries of the system. The maximal radius is given by the solution of these equations at $\tau = 0$ as
\begin{equation}
 r_{max} = r(\tau = 0) = GM + \sqrt{G^2M^2 + E_m^2}
 \label{rmax}
\end{equation}
matching with the results obtained in \cite{Larsen98}.
The contraction of the string continues till it reaches the event horizon, giving us the minimum radius as
\begin{equation}
r_{min} = 2GM
\label{rmin}
\end{equation}

Thus, equations (\ref{rmax}) and (\ref{rmin}) give the radial limits on the size of the loop, heavily dependent on the mass of the black hole.  Furthermore, the black hole mass also governs the contraction (decay) time of the string, from $r_{max}$ to $r_{min}$, calculated using  (\ref{tdot}),
   
\begin{equation}
t_{\rm BH}(\tau) = E_m \Delta\tau + 2GM \log \left| \frac{ \tan(\Delta\tau/2) +  \epsilon }{ \tan(\Delta\tau/2) -  \epsilon} \right|
\label{decaytime}
\end{equation}
 where $\epsilon = (\sqrt{G^2M^2 + E^2}-GM)/E_m$ and the world-sheet time $\Delta \tau$ is given by
\begin{equation}
\Delta \tau = cos^{-1}\left(\frac{GM}{\sqrt{G^2 M^2 + E_m^2}}\right)
\label{deltatau}
\end{equation}

\begin{figure}[h!]
	\centering
	\includegraphics[width=0.7\linewidth]{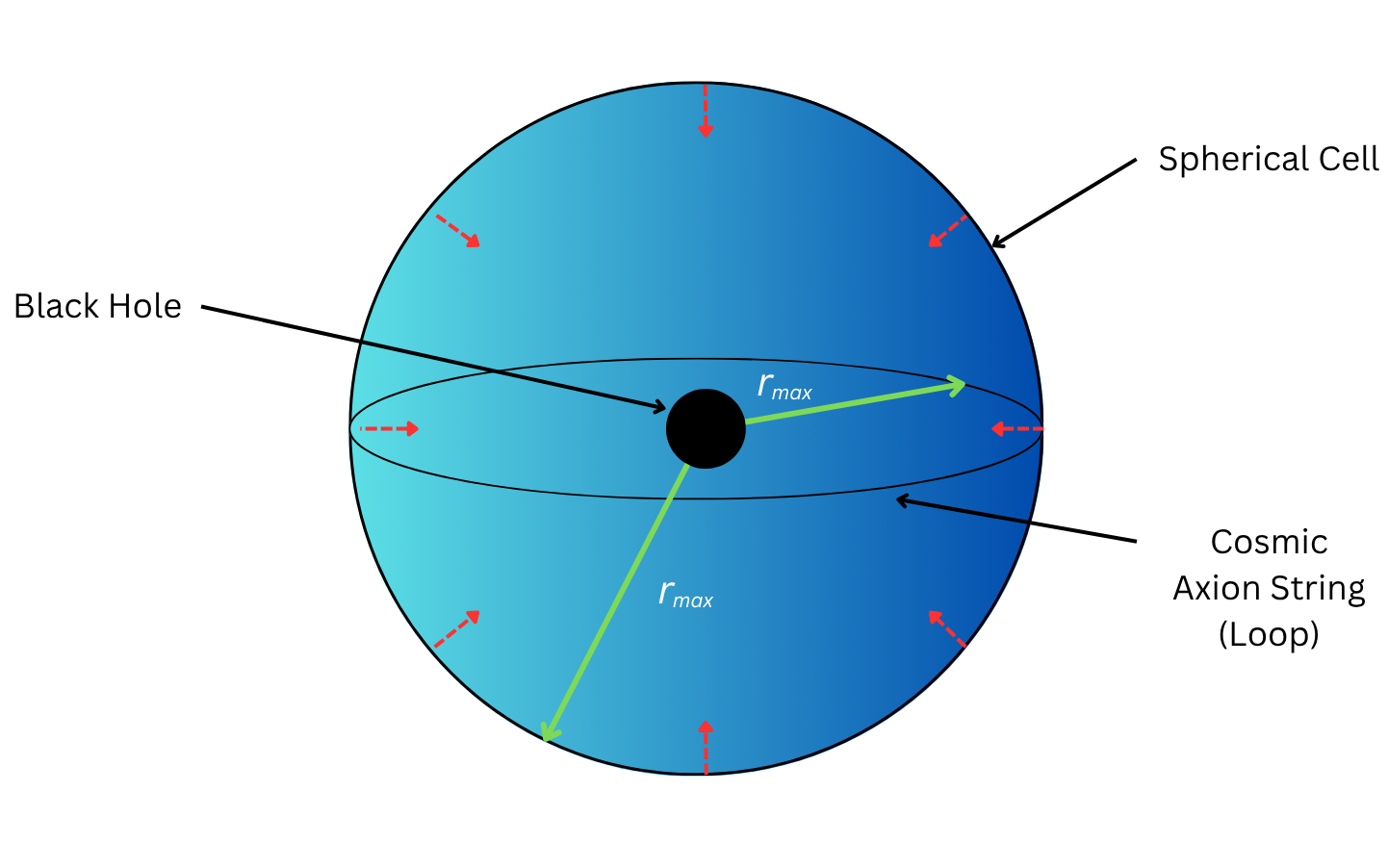}
	\caption{Depiction of the spherical cell considered for the cosmic string-black hole system}
	\label{fig1}
\end{figure}

As the string contracts and forms kinks under the influence of the black hole, it dissipates energy, primarily in the form of axions. To quantify this radiated energy, we consider a cell of spherical volume having radius $r$ and one string loop of the same radius inside it for spatial symmetry as illustrated in Fig. \ref{fig1} (Image not to scale). The red arrows indicate the direction of the contraction of the cell with the contracting string.
It can be easily calculated that the energy inside the cell is 
\begin{equation}
E(r) =  4\pi^2 f_a^2 r \ln(2 \pi r/\delta) 
\label{Energy1}
\end{equation}
As the string loop contracts, kinks appear causing the emission by the string to maintain the energy inside the contracting cell \cite{Davis86},\cite{Vilenkin00}. Hence, the radiated energy evaluated (at some new radius $r' < r$) as  
\begin{align}
\Delta E  &= E(r) - E(r')\nonumber \\
		 &= 4\pi^2f_a^2  \left(r\ln(2\pi r/\delta) - r'\ln(2\pi r'/\delta) \right)\nonumber \\
		 &= 4\pi^2f_a^2 \Delta(r\ln(2 \pi r/ \delta))
\label{deltaE}
\end{align}

Now, by applying limits on radius from (\ref{rmax}) and (\ref{rmin}) in (\ref{deltaE}), the energy radiated by the string during contraction has been calculated as, 
\begin{align}
\Delta E =\;& 4\pi^2 f_a^2 \bigg((GM + \sqrt{G^2 M^2 + E_m^2}) 
\ln\left(\frac{2\pi (GM + \sqrt{G^2 M^2 + E_m^2})}{\delta}\right) \nonumber \\
& \quad - 2GM \ln\left(\frac{4\pi GM}{\delta}\right) \bigg)
\label{Erad}
\end{align}
and the energy emission rate can be shown to be $\Delta E/\Delta \tau$. Further, it can also be established that the energy density inside the cell increases as
\begin{equation}
\Delta \rho (r) = \frac{\Delta E}{\frac{4}{3} \pi r^3} = \frac{3 \pi f_a^2}{r^3} \Delta(r\ln(2 \pi/ \delta))
\label{density}
\end{equation}

Interestingly, we have found that the radiated energy and the rate of radiation remains nearly constant in this system with respect to the fixed mass of the black hole, the initial (maximum) energy and decay constant of the string. Given that the radiation rate remains nearly constant for a fixed black hole mass, this mechanism points toward a stable, continuous background of axion emission.\\

\section{Numerical analysis}\label{analysis}

\subsection{Energy Radiation}
With the equations for the radiated energy established in the previous section, it is important to get an estimate on the values of $\Delta E$. These values will help us in understanding the importance of this mechanism and the significance of its contribution to the energy of the universe. 

\begin{figure}[ht]
\centering
\includegraphics[width=0.7\linewidth]{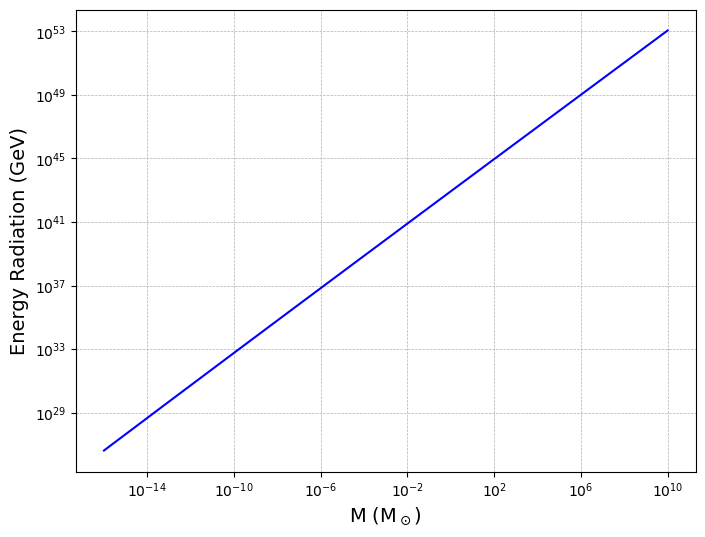 }
\caption{$\Delta E$ for varying black hole mass $(M)$}
\label{fig2}
\end{figure}

\begin{figure}[ht]
\centering
\includegraphics[width = 0.7\linewidth] {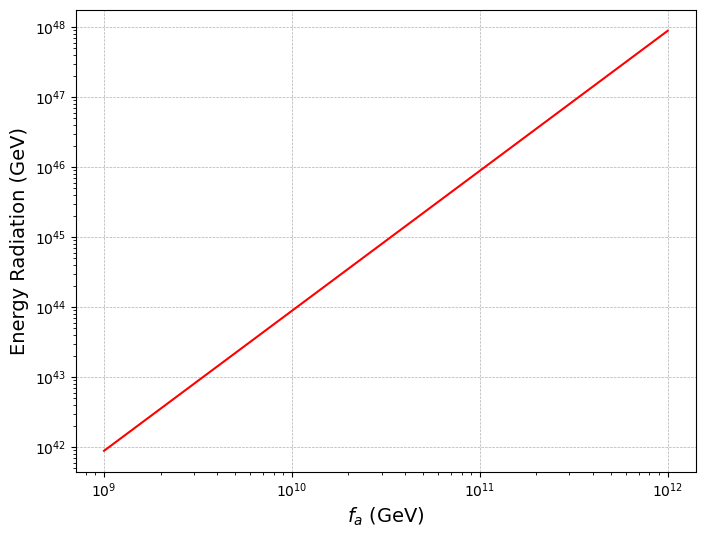 }
\caption{$\Delta E$ for varying decay constant ($f_a$)}
\label{fig3}
\end{figure}

In this regard, we simulate $\Delta E$ over a mass range starting from $10^{-16} M_\odot$ of primordial black holes (PBHs) upto the largest $10^{10} M_\odot$ of supermassive black holes (SMBHs) for a decay constant $f_a = 10^{10}$ GeV depicted in Fig. \ref{fig2} (both axes are in logscale). Similarly, Fig. \ref{fig3} further examines energy dependence on decay constant in the range $f_a \in (10^9, 10^{12})$ GeV \cite{Preskill83} by taking $M=10 M_\odot$. The string core radius corresponds to $\sim 1/f_a$. Simulations and plots were implemented using Python’s Matplotlib library.


\begin{figure}[h!]
\centering
\includegraphics[width = 0.7\linewidth] {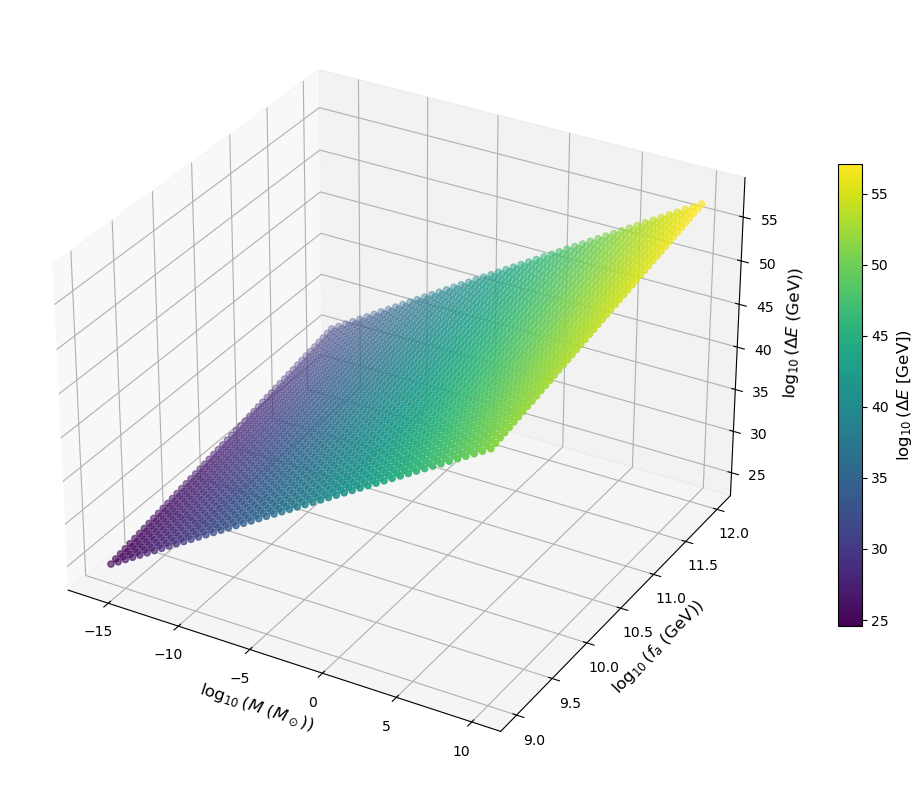 }
\caption{$\Delta E$ for varying black hole mass $(M)$ and decay constant $f_a$}
\label{fig4}
\end{figure}

Fig. \ref{fig2} and Fig. \ref{fig3} demonstrate the linear relationship between energy gain and mass and decay constant as expected from (\ref{Erad}) . It is interesting to note that even for the smallest black holes, the energy radiated is as high as $\sim \mathcal{O}(10^{27})$ GeV. It is emphasized here that the order of the energy radiated is affected when a different decay constant or black hole mass is considered, but the plot's behavior remains unchanged. Fig.\ref{fig4} illustrates the same behavior by displaying the combined effect of variations in the decay constant and the black hole's mass as in (\ref{deltatau}).\\

\begin{figure}[h!]
\centering
\includegraphics[width = 0.7\linewidth] {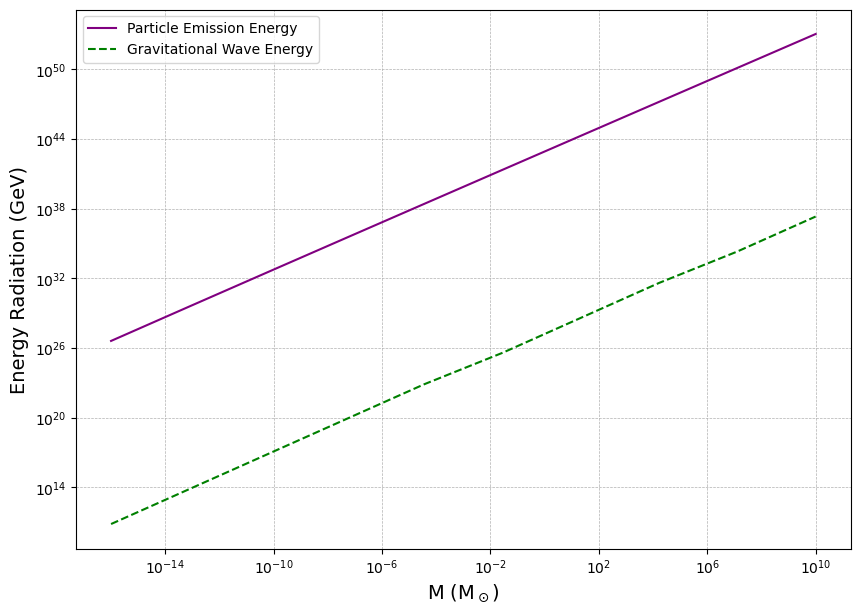 }
\caption{$\Delta E$ corresponding to the GWs (green dotted line) and particle emission (purple line) for varying mass}
\label{fig5}
\end{figure}

The energy calculated and plotted in Fig. \ref{fig2}, Fig. \ref{fig3} and Fig. \ref{fig4} illustrates the total energy released by this mechanism. String loop decay primarily dissipates energy through two channels: particle emission, specifically axions in this context, and gravitational radiation. However, recent findings \cite{Baeza2024} indicate that power radiated GW ($P_{\rm GW}$) is considerably suppressed in comparison to power radiated in particle emission ($P_{\varphi}$)  quantified as
\begin{equation}
\frac{P_{\rm GW}}{P_{\varphi}} \approx \mathcal{O}(10) \left( \frac{v}{m_{\rm p}} \right)^2 \ll 1.
\label{energyratio}
\end{equation}
where $v$ is the vacuum expectation value associated to string formation and $m_{\rm p}$ is the Planck mass. Applying this relation (\ref{energyratio}) in our framework allows a clear separation of the energy into gravitational wave and particle radiation  This distinction is depicted in Fig. \ref{fig5}, demonstrating the dominance of particle radiation.

\subsection{Decay Time}\label{timeanalysis}
\begin{figure}[h!]
\centering
\includegraphics[width = 0.7\linewidth] {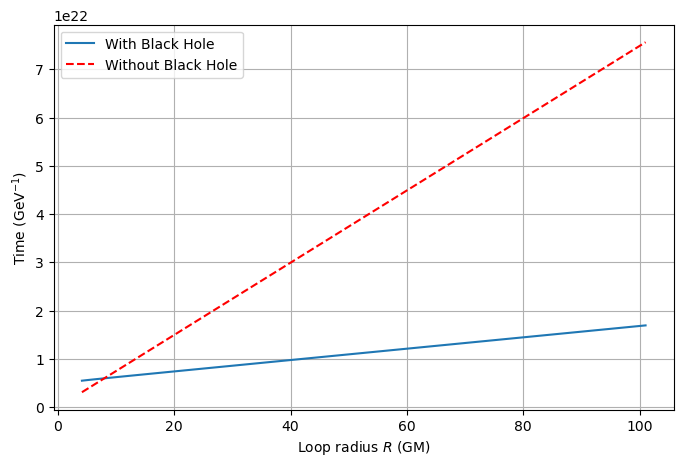 }
\caption{Decay time with (blue line) and without (red dashed line) a black hole for varying black hole mass}
\label{fig6}
\end{figure}

\begin{figure}[h!]
\centering
\includegraphics[width = 0.7\linewidth] {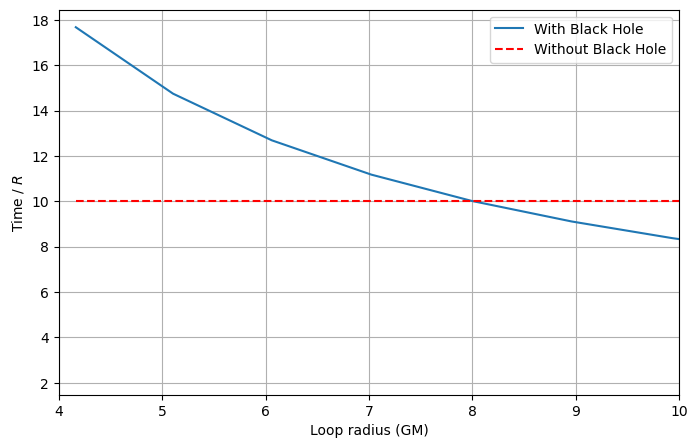 }
\caption{Decay time/loop radius with (blue line) and without (red dashed line) a black hole for varying black hole mass depicing the increased decay time close to the black hole}
\label{fig7}
\end{figure}

The decay time for our black hole - cosmic string system is estimated using (\ref{decaytime}) and (\ref{deltatau}) for an observer at infinity.  In the absence of black hole however, the decay time depends linearly on the loop length $L$ as described by \cite{Saurabh20}
\begin{equation}
t_{\rm free} \sim 1.4 L
\label{decayfree}
\end{equation} 
measured in string's rest frame. Incorporating the Lorentz boost factor of 1.25, Fig. \ref{fig6} compares decay times with and without the central black hole, clearly showing accelerated decay induced by the black hole’s gravitational field. An additional noteworthy observation is the higher decay time up to a loop radius of 8 $GM$, illustrated in Fig. \ref{fig7}, which arises from the gravitational time dilation caused by the black hole. This effect has significant implications for potential observational signatures of the system.

\section{Results and discussions}\label{results}

We explore the dynamical impact of black holes on the cosmic string energy by calculating the total axion radiation for various mass scales and decay constants. This analysis allows us to quantify how the strong-field environment of the black hole triggers the string’s contraction and determines the magnitude of the resulting energy radiation. Our simulations cover a wide range of black hole masses from primordial black holes (\(\sim 10^{-16} M_\odot\)) to supermassive black holes (\(\sim 10^{10} M_\odot\)) with typical axion decay constants between \(10^9\) and \(10^{12}\) GeV.

The total radiated energy ranges from approximately \(10^{27}\) GeV for the smallest black holes to as high as \(10^{53}\) GeV for supermassive black holes, indicating a substantial release of axion radiation in these systems. Given the enormity of this energy release, axion strings colliding with black holes may be a non-trivial source of high-energy particle injection in galactic environments or the early universe. Such vast emission implies that these systems might have a substantial impact on the local axion energy density, possibly influencing the thermal history of the surrounding plasma or adding to the diffuse axion background. 

Significantly, our analysis reveals that the decay times of cosmic string loops are considerably shortened by the gravitational influence of the central black hole. This accelerated decay serves as a notable effect that could impact both the dynamics of axion emission and possible observational signals, especially through gravitational wave co-emission. Gravitational wave emission, although suppressed relative to axion particle emission, is non-negligible and represents a promising channel for observational signatures of this mechanism. \\

To provide a definitive assessment of the axion radiation spectrum, further investigation is required into the time-dependent interplay between black hole growth and string contraction. Accounting for the evolution of the central mass and the cumulative effects of loop decay over cosmic time scales will be essential for predicting the exact energy distribution of the radiated axions. \\

Beyond theoretical modeling, these dynamics have direct implications for future observational surveys. Specifically, it has been shown that \cite{Agashe23} future gamma-ray telescopes such as e-ASTROGAM could detect ALPs emitted by PBHs if these particles decay to photons before reaching Earth, producing distinctive double-peaked spectral signatures compared to Standard Model predictions. This method of photon coupling could similarly be applied in detecting the axions radiated due to string contraction in black hole spacetime. Additionally, as PBHs grow into SMBHs, the parameters determining loop contraction become dynamic, further affecting the total energy radiated. These factors highlight that the scenario described does not capture the complete picture and must be refined to account for evolving nature of PBH mass and string loop dynamics.\\

\section{Conclusion}
\label{Conclusion}

This study investigates a Schwarzschild black hole encircled by a circular cosmic axion string, where  the string’s equation of motion and extremal values for loop size have been analysed. Our results demonstrate that as the string contracts, the resulting formation of kinks serves as a powerful engine for axion radiation, with an energy output sensitive to the black hole mass, the loop's geometric scale, and the axion decay constant. 

Numerical results across the mass spectrum—ranging from Primordial Black Holes (PBHs) to Supermassive Black Holes (SMBHs)—quantify a significant axion and GW radiation into the surrounding environment. The presence of a central mass not only accelerates the decay chronology of these loops but also modifies the total power spectrum of the emitted axions. These findings point toward a promising multi-messenger observational landscape, where the characteristic high-energy particle background is accompanied by distinct gravitational wave signatures.

Future work will need to incorporate more complex string configurations and evolving system parameters to rigorously test the viability of axion emission from black hole-cosmic string systems as a dominant dark matter genesis mechanism.

\bibliography{Citations}

@article{Baeza2024,
  title = {Gravitational wave emission from a cosmic string loop: Global case},
  author = {Baeza-Ballesteros, Jorge and Copeland, Edmund J. and Figueroa, Daniel G. and Lizarraga, Joanes},
  journal = {Phys. Rev. D},
  volume = {110},
  issue = {4},
  pages = {043522},
  numpages = {12},
  year = {2024},
  month = {Aug},
  publisher = {American Physical Society},
  doi = {10.1103/PhysRevD.110.043522},
  url = {https://link.aps.org/doi/10.1103/PhysRevD.110.043522}
}

@article{Swamyprd,
  title = {Blandford-Znajek jets and the total angular momentum evolution of a black hole connected to a cosmic string},
  author = {Swamy, Ishan and Singh, Deobrat},
  journal = {Phys. Rev. D},
  pages = {--},
  year = {2025},
  month = {Aug},
  publisher = {American Physical Society},
  doi = {10.1103/58tc-v253},
  url = {https://link.aps.org/doi/10.1103/58tc-v253}}

@article{Agashe23,
  title = {Detecting axionlike particles with primordial black holes},
  author = {Agashe, Kaustubh and Chang, Jae Hyeok and Clark, Steven J. and Dutta, Bhaskar and Tsai, Yuhsin and Xu, Tao},
  journal = {Phys. Rev. D},
  volume = {108},
  issue = {2},
  pages = {023014},
  numpages = {13},
  year = {2023},
  month = {Jul},
  publisher = {American Physical Society},
  doi = {10.1103/PhysRevD.108.023014},
  url = {https://link.aps.org/doi/10.1103/PhysRevD.108.023014}
}

@article{Preskill83,
title = {Cosmology of the invisible axion},
journal = {Physics Letters B},
volume = {120},
number = {1},
pages = {127-132},
year = {1983},
issn = {0370-2693},
doi = {https://doi.org/10.1016/0370-2693(83)90637-8},
url = {https://www.sciencedirect.com/science/article/pii/0370269383906378},
author = {John Preskill and Mark B. Wise and Frank Wilczek}
}

@article{Larsen98,
    author = "Larsen, Arne L. and Nicolaidis, Argyris",
    title = "{String spreading on black hole horizon}",
    eprint = "gr-qc/9812059",
    archivePrefix = "arXiv",
    doi = "10.1103/PhysRevD.60.024012",
    journal = "Phys. Rev. D",
    volume = "60",
    pages = "024012",
    year = "1999"
}

@article{Davis86,
    author = "Davis, Richard Lynn",
    title = "{Cosmic Axions from Cosmic Strings}",
    reportNumber = "SLAC-PUB-3895",
    doi = "10.1016/0370-2693(86)90300-X",
    journal = "Phys. Lett. B",
    volume = "180",
    pages = "225--230",
    year = "1986"
}

@article{Turok84,
title = {Grand unified strings and galaxy formation},
journal = {Nuclear Physics B},
volume = {242},
number = {2},
pages = {520-541},
year = {1984},
issn = {0550-3213},
doi = {https://doi.org/10.1016/0550-3213(84)90407-3},
url = {https://www.sciencedirect.com/science/article/pii/0550321384904073},
author = {Neil Turok}
}

@article{Vachaspati85,
  title = {Gravitational radiation from cosmic strings},
  author = {Vachaspati, Tanmay and Vilenkin, Alexander},
  journal = {Phys. Rev. D},
  volume = {31},
  issue = {12},
  pages = {3052--3058},
  numpages = {0},
  year = {1985},
  month = {Jun},
  publisher = {American Physical Society},
  doi = {10.1103/PhysRevD.31.3052},
  url = {https://link.aps.org/doi/10.1103/PhysRevD.31.3052}
}

@article{Vilenkin87,
  title = {Radiation of Goldstone bosons from cosmic strings},
  author = {Vilenkin, Alexander and Vachaspati, Tanmay},
  journal = {Phys. Rev. D},
  volume = {35},
  issue = {4},
  pages = {1138--1140},
  numpages = {0},
  year = {1987},
  month = {Feb},
  publisher = {American Physical Society},
  doi = {10.1103/PhysRevD.35.1138},
  url = {https://link.aps.org/doi/10.1103/PhysRevD.35.1138}
}

@article{Matsunami19,
  title = {Decay of Cosmic String Loops due to Particle Radiation},
  author = {Matsunami, Daiju and Pogosian, Levon and Saurabh, Ayush and Vachaspati, Tanmay},
  journal = {Phys. Rev. Lett.},
  volume = {122},
  issue = {20},
  pages = {201301},
  numpages = {5},
  year = {2019},
  month = {May},
  publisher = {American Physical Society},
  doi = {10.1103/PhysRevLett.122.201301},
  url = {https://link.aps.org/doi/10.1103/PhysRevLett.122.201301}
}

@article{Peccei77,
  title = {$\mathrm{CP}$ Conservation in the Presence of Pseudoparticles},
  author = {Peccei, R. D. and Quinn, Helen R.},
  journal = {Phys. Rev. Lett.},
  volume = {38},
  issue = {25},
  pages = {1440--1443},
  numpages = {0},
  year = {1977},
  month = {Jun},
  publisher = {American Physical Society},
  doi = {10.1103/PhysRevLett.38.1440},
  url = {https://link.aps.org/doi/10.1103/PhysRevLett.38.1440}
}

@article{Weinberg78,
  title = {A New Light Boson?},
  author = {Weinberg, Steven},
  journal = {Phys. Rev. Lett.},
  volume = {40},
  issue = {4},
  pages = {223--226},
  numpages = {0},
  year = {1978},
  month = {Jan},
  publisher = {American Physical Society},
  doi = {10.1103/PhysRevLett.40.223},
  url = {https://link.aps.org/doi/10.1103/PhysRevLett.40.223}
}

@article{Abbott83,
  author = {L. F. Abbott and P. Sikivie},
  title = {A cosmological bound on the invisible axion},
  journal = {Physics Letters B},
  volume = {120},
  pages = {133--136},
  year = {1983},
  doi = {10.1016/0370-2693(83)90638-X}
}

@article{Dine83,
  author = {Michael Dine and Willy Fischler},
  title = {The not-so-harmless axion},
  journal = {Physics Letters B},
  volume = {120},
  pages = {137--141},
  year = {1983},
  doi = {10.1016/0370-2693(83)90639-1}
}

@misc{Vilenkin00,
  author    = {Vilenkin, A. and Shellard, E. P. S.},
  title     = {Cosmic Strings and Other Topological Defects},
  publisher = {Cambridge University Press},
  year      = {2000}
}

@article{Saurabh20,
    author = "Saurabh, Ayush and Vachaspati, Tanmay and Pogosian, Levon",
    title = "{Decay of Cosmic Global String Loops}",
    doi = "10.1103/PhysRevD.101.083522",
    journal = "Phys. Rev. D",
    volume = "101",
    number = "8",
    pages = "083522",
    year = "2020"
}

@article{Vilenkin86,
    author = "Vilenkin, Alexander and Vachaspati, Tanmay",
    title = "{Radiation of Goldstone Bosons From Cosmic Strings}",
    reportNumber = "TUTP-86-16",
    doi = "10.1103/PhysRevD.35.1138",
    journal = "Phys. Rev. D",
    volume = "35",
    pages = "1138",
    year = "1987"
}

@article{Churilova24,
    author = "Churilova, Mariia and Kolo\v{s}, Martin and Stuchl\'\i{}k, Zden\v{e}k",
    title = "{String loop vibration around Schwarzschild black hole}",
    doi = "10.1140/epjc/s10052-023-12367-0",
    journal = "Eur. Phys. J. C",
    volume = "84",
    number = "1",
    pages = "25",
    year = "2024"
}

@article{Kinoshita16,
    author = {S. Kinoshita and T. Igata and K. Tanabe},
    title = {Extraction of {R}otational {E}nergy from a {B}lack {H}ole},
    journal = {Phys. Rev. D},
    volume = {94, {124039}},
    year = {2016}}

@article{Xing21,
    author = {H. Xing and Y. Levin and A. Gruzinov and A. Vilenkin},
    title = {Spinning black holes as cosmic string factories},
    journal = {Phys. Rev. D},
    volume = {103, {083019}},
    year = {2021}}

@article{Larsen94,
doi = {10.1088/0264-9381/11/5/008},
year = {1994},
publisher = {},
volume = {11},
number = {5},
pages = {1201},
author = {A L Larsen},
title = {Chaotic string-capture by black hole},
journal = {Classical and Quantum Gravity},
}

@article{Frolov99,
doi = {10.1088/0264-9381/16/11/316},
year = {1999},
publisher = {},
volume = {16},
number = {11},
pages = {3717},
author = {Andrei V Frolov and Arne L Larsen},
title = {Chaotic 
scattering and capture of strings by 
a black hole},
journal = {Classical and Quantum Gravity},
}

@article{Swamy25,
title = {Impact on orbital period of X-ray binary systems attached to a cosmic string},
journal = {New Astronomy},
volume = {118},
pages = {102378},
year = {2025},
issn = {1384-1076},
doi = {https://doi.org/10.1016/j.newast.2025.102378},
author = {Ishan Swamy and Deobrat Singh},
}

@article{Singh25,
    author = "Swamy, I. and Singh, D.",
    title = "{Investigating the interaction of a cosmic string with an accreting black hole}",
    doi = "10.1088/1402-4896/adce49",
    journal = "Phys. Scripta",
    volume = "100",
    number = "6",
    pages = "065008",
    year = "2025"
}

@article{Vilenkin18,
    author = "Vilenkin, Alexander and Levin, Yuri and Gruzinov, Andrei",
    title = "{Cosmic strings and primordial black holes}",
    doi = "10.1088/1475-7516/2018/11/008",
    journal = "JCAP",
    volume = "11",
    pages = "008",
    year = "2018"
}

@article{Berezhiani91,
  author  = {Z. G. Berezhiani and M. Y. Khlopov},
  title   = {Cosmology of spontaneously broken gauge family symmetry with axion solution of strong CP-problem},
  journal = {Zeitschrift für Physik C Particles and Fields},
  volume  = {49},
  number  = {1},
  pages   = {73--78},
  year    = {1991}
}

@article{parth24,
  author  = {P. Bambhaniya and O. Trivedi and I. Dymnikova and P. S. Joshi and M. Khlopov},
  title   = {On the interactions of black holes and cosmic strings},
  journal = {Physics of the Dark Universe},
  volume  = {46},
  pages   = {101553},
  year    = {2024}
}

@article{Khlopov99,
  author  = {M. Y. Khlopov and A. S. Sakharov and D. D. Sokoloff},
  title   = {The nonlinear modulation of the density distribution in standard axionic {CDM} and its cosmological impact},
  journal = {Nuclear Physics B - Proceedings Supplements},
  volume  = {72},
  pages   = {105--109},
  year    = {1999}
}

\end{document}